# Structural evidence for ultrafast polarization rotation in ferroelectric/dielectric superlattice nanodomains

Hyeon Jun Lee[1], Youngjun Ahn[1], Samuel D. Marks[1], Eric C. Landahl[2], Shihao Zhuang[1], M. Humed Yusuf[3], Matthew Dawber[3], Jun Young Lee[4], Tae Yeon Kim[4], Sanjith Unithrattil[4], Sae Hwan Chun[5], Sunam Kim[5], Intae Eom[5], Sang-Yeon Park[5], Kyung Sook Kim[5], Sooheyong Lee[6,7], Ji Young Jo[4], Jiamian Hu[1], and Paul G. Evans[1,*]

[1]Department of Materials Science and Engineering, University of Wisconsin-Madison, Madison, Wisconsin 53706, USA

[2]Department of Physics, DePaul University, Chicago, Illinois 60614, USA

[3]Department of Physics and Astronomy, Stony Brook University, Stony Brook, New York 11794, USA

[4]School of Materials Science and Engineering, Gwangju Institute of Science and Technology, Gwangju 61005, South Korea

[5]Pohang Accelerator Laboratory, Pohang, Gyeongbuk 37673, South Korea

[6]Korea Research Institute of Standards and Science, Daejeon 34113, South Korea

[7]Department of Nano Science, University of Science and Technology, Daejeon 34113, South Korea

[*]pgevans@wisc.edu

Weakly coupled ferroelectric/dielectric superlattice thin film heterostructures exhibit complex nanoscale polarization configurations that arise from a balance of competing electrostatic, elastic, and domain-wall contributions to the free energy. A key feature of these configurations is that the polarization can locally have a significant component that is not along the thin-film surface normal direction, with an overall configuration maintaining zero net in-plane polarization. $PbTiO_3$/$SrTiO_3$ thin-film superlattice heterostructures on a conducting $SrRuO_3$ bottom electrode on $SrTiO_3$ have a room-temperature stripe nanodomain pattern with nanometer-scale lateral period. Ultrafast time-resolved x-ray free electron laser diffraction and scattering experiments reveal that above-bandgap optical pulses induce rapidly propagating acoustic pulses and a perturbation of the domain diffuse scattering intensity arising from the nanoscale stripe domain configuration. With 400 nm optical excitation, two separate acoustic pulses are observed: a high-amplitude pulse resulting from strong optical absorption in the bottom electrode and a weaker pulse arising from the depolarization field screening effect due to absorption directly within the superlattice. The picosecond scale variation of the nanodomain diffuse scattering intensity is consistent with a larger polarization change than would be expected due to the polarization-tetragonality coupling of uniformly polarized ferroelectrics. The polarization change is consistent instead with polarization rotation facilitated by the reorientation of the in-plane component of the polarization at the domain boundaries of the striped polarization structure. The complex steady-state configuration within these ferroelectric heterostructures leads to polarization rotation phenomena that have been previously available only through the selection of bulk crystal composition.

# I. INTRODUCTION

Epitaxial ferroelectric/dielectric superlattices (SLs) are highly engineered nanomaterials that can exhibit complex configurations of their electric polarization, ranging from nanoscale striped domains to chiral vortices and Skyrmions [1-3]. The absorption of optical radiation with above-bandgap photon energy leads to a series of effects that ultimately shift the boundary conditions governing the formation of these configurations leading to a range of phenomena that are uniquely enabled by the nanoscale spatial variation of the polarization. In the absence of complex nanoscale polarization patterns, the response of ferroelectrics to above-bandgap radiation is relatively straightforward: photoinduced depolarization field screening produces lattice expansion on a picosecond timescale set by acoustic phenomena [4,5]. The complexity of the polarization configuration within the nanodomains, however, changes the optical response at all length scales ranging from the atomic scale of the ferroelectric polarization to the mesoscopic domain structure because the polarization can be distorted by a local change in orientation [6]. Here we show that the nanoscale domain pattern has a dramatic effect on atomic-scale polarization dynamics. We report the discovery of an unusual relationship between polarization and lattice distortion in ferroelectric/dielectric $PbTiO_3$/$SrTiO_3$ (PTO/STO) SLs that is fundamentally linked to the complex polarization configuration within these SLs. The change in polarization arising from optically induced out-of-plane strain is stronger than the relationship predicted by the longstanding phenomenological thermodynamic theory applicable to uniform ferroelectrics. Polarization rotation within nanodomains mediates a new relationship between polarization and strain, producing optically induced polarization increases that can have a magnitude as large as 22%. Ultimately, this optically induced polarization rotation can provide a mechanism for ultrafast modulation of the polarization and of effects linked to electronic properties such as nonlinear-optical coefficients.

The spontaneous polarization of ferroelectrics has a strong coupling to the dimensions of the crystallographic unit cell. In the case where the lattice distortion is along the same axis as the polarization (e.g., in the case of tetragonal ferroelectrics such as PTO in the form of bulk crystals or uniform-composition thin films), the polarization is accurately described using phenomenological thermodynamic models based on the Landau-Ginzburg-Devonshire (LGD) theory [7,8]. In the LGD description, the out-of-plane polarization ($P_z$) of a tetragonal ferroelectric thin film is proportional to the square root of the out-of-plane tetragonality [8]. This square-root dependence has been experimentally verified in uniformly polarized bulk and thin-film ferroelectrics [9,10]. In PTO/STO SLs, however, bound charges at the surface and interface between the PTO and STO layers create an electric field opposite to the polarization in PTO layer, the depolarization field. This depolarization field leads to the formation of an intricate nanoscale polarization distribution in which, crucially, there is a pattern of in-plane polarization components with zero net total polarization, in addition to the usual out-of-plane polarization [3,11-14].

The local polarization configuration of ferroelectrics is closely linked to the distortion of their crystal structures, making structural studies particularly valuable in understanding nanoscale ferroelectricity. Direct insight into the nanoscale organization of the polarization in PTO/STO SLs can be obtained by studying the reciprocal-space distribution of the diffuse x-ray scattering intensity arising from the nanoscale polarization configuration. Here, we apply ultrafast x-ray scattering measurements probing ferroelectric nanodomain diffuse scattering to determine the perturbation of the ferroelectric polarization induced by ultrafast optical excitation, as shown in Fig. 1(a).

Beyond the description given by the LGD expressions, previous studies have revealed that the polarization and structural distortion can evolve separately. The tetragonality and the polarization can be individually probed and can have different picosecond-timescale dynamics [15]. Ultrafast powder x-ray diffraction experiments separately track the dynamics of the polarization and the time-dependent variation of the lattice parameters of oxide and molecular ferroelectrics [16,17]. The ultrafast optically induced changes in polarization can include transient reversals in the sign of the nonlinear optical coefficient $\chi^{(2)}$, corresponding to a reversal of polarization [18]. Optically driven effects can lead to transformations among energetically similar structural phases or among ferroelastic variants of a single phase [19,20]. Structural distortion can lead to dynamical changes in the Born effective charge and thus the polarization [21].

Ferroelectrics in which the polarization is not collinear with an applied electric field exhibit field- or stress-induced polarization rotation, through which the electric field (or elastic deformation) can reorient the polarization. In general, this effect shifts the direction of the polarization so that it points closer to the direction set by the electric field or the expansion and includes a complex series of geometric pathways along which the polarization evolves [6,22]. A common feature of these pathways is that the polarization depends strongly on the overall expansion of the out-of-plane lattice. Polarization rotation has important functional consequences, for example, leading the local susceptibility in regions with in-plane polarization to be higher than at the center of domains [11,23]. A schematic of how the polarization rotation mechanism can occur within the nanodomain polarization configuration of the SLs is shown in Fig. 1(b). This mechanism is discussed and tested in detail below.

## II. STRUCTURAL DYNAMICS FOLLOWING OPTICAL EXCITATION

Optical excitation leads to three separate sources of elastic distortion within the SL. First, lattice expansion results from depolarization field screening. In general, depolarization field screening arises from the absorption of optical radiation in ferroelectric thin films, inducing stress that leads to lattice expansion. The depolarization field screening effect has been extensively studied in ferroelectric thin films with uniform composition, including PTO and $BiFeO_3$ [4,5]. Here we extend the discussion of the depolarization field screening effect to a PTO/STO ferroelectric/dielectric SL. The depolarization-field screening expansion can be unambiguously distinguished from heating of the SL because PTO-based ferroelectrics, including the PTO/STO SL, exhibit a thermal contraction rather than the depolarization field screening-induced expansion [5,11,24]. Other systems in which there is a competition between ferroic and elastic effects can also exhibit a response arising from long-range order that has a different sign than the effects of thermal expansion, as for example, in FePt thin films under some elastic conditions [25]. Second, an acoustic pulse is produced due to optical absorption in the conducting oxide bottom electrode on which our particular heterostructure was grown. Optical absorption in the electrode leads to rapid heating, thermal expansion, and the generation of a photoacoustic pulse [26]. The acoustic pulses generated through these two mechanisms propagate through the thickness of the epitaxial heterostructure at the longitudinal acoustic sound velocity [27,28]. An additional contribution arising from the relatively slow heat transport via conduction from the SRO electrode to the SLs leads to heating of the SL in the region near the SL/SRO interface over timescales of tens of ps and to a lattice contraction due to the negative coefficient of thermal expansion in the SL [24,29].

Optically induced lattice and polarization phenomena in the SL were probed in diffraction experiments at the Pohang Accelerator Laboratory X-ray Free-electron Laser (PAL-XFEL) [30]. The time resolution of these experiments was 50 fs, faster than the timescales of structural expansion and domain motion. The nanodomain pattern produces a ring of diffuse scattering with an in-plane reciprocal-space radius and an intensity 2-3 orders of magnitude lower than the Bragg reflections. Intensity maxima arising from domain diffuse scattering appear at $Q_x = \pm 0.054$ Å$^{-1}$ in the equilibrium configuration diffraction pattern shown in Fig. 1(c), which was acquired at a time delay in which x-rays arrive before the optical pulse. At a fixed x-ray incident angle each detector image measured the intensity in a section of reciprocal space including two intersections with the ring of domain diffuse scattering intensity. The out-of-plane lattice distortion of the SL was determined by measuring the intensity variation along $Q_z$ near the PTO/STO SL 002 Bragg reflection. Further details of the ultrafast x-ray diffraction measurement are given in section V.A.

The repeating unit of the PTO/STO SL consisted of 8 unit cells (u. c.) of PTO and 3 u. c. of STO. The SL thin film had a total thickness $d=100$ nm on a 20 nm-thick SrRuO$_3$ (SRO) bottom electrode on an STO substrate. Under these strain and layer-thickness conditions the equilibrium polarization configuration is a striped nanodomain pattern with a lateral period of 11.6 nm [31]. The SL heterostructure was excited with 50-fs-duration optical pulses with wavelength 400 nm and a repetition rate matching the 30 Hz x-ray pulse repetition rate. The scattering experiments were conducted at 395 K in order to eliminate artifacts associated with slow room-temperature charge-trapping dynamics [32].

The two optical excitation mechanisms described above lead to out-of-plane strain resulting in a change in the average lattice parameter of nanodomains and to a change in the volume-averaged $P_z$ in each stripe of the nanodomains. The photoinduced distortion can be approximately separated into time regimes defined by the timescale of acoustic propagation in which different effects are the dominant sources of strain. These effects are illustrated in Fig. 1(d):

(1) the propagation of transient acoustic pulses over tens of ps for films with thicknesses on the order of 100 nm and (2) the depolarization field screening effect due to excited charge carriers that begins immediately upon optical absorption but persists for a longer duration and becomes the dominant effect only after the acoustic wave propagation is complete. Note that although the STO polarization is already low in comparison with the PTO component, the depolarization field screening leads to a decrease in polarization following excitation. The approximate time-separation of these phenomena allows their effects on the polarization to be evaluated independently.

Acoustic pulses propagate through the overall thickness of the SL with a transit time $\tau=d/v$, where $v=3700$ m s$^{-1}$ is the longitudinal sound velocity in the SL. The value of $v$ was precisely determined from the time-resolved diffraction measurements [33], as described in section V.B. The strain distribution evolves through acoustic phenomena including acoustic reflection at the free surface and reflection and transmission at the substrate-film interface.

The time dependence of the domain diffuse scattering intensity is shown in Fig. 1(e) for an optical pump fluence of 1.17 mJ cm$^{-2}$. Optical excitation is followed by a variation of the domain diffuse scattering intensity, which provides key insight into the polarization within the nanodomains because the intensity is closely linked to the magnitude and spatial distribution of the polarization $P_z$ in the domain pattern. This variation is discussed extensively below.

The response to the optical pulses also induces a shift along $Q_z$ of the maximum intensity of the domain diffuse scattering by $\Delta Q_z$, as shown in Fig. 1(f). The overall time dependence of $\Delta Q_z$ must be described before the more detailed discussion of the origin of the polarization changes. We thus focus first on $\Delta Q_z$, which has a complex time dependence arising from the rapidly varying strain during the acoustic pulses. An initial contraction of the average lattice parameter occurs immediately following the optical excitation, yielding an initially positive $\Delta Q_z$. Acoustic simulations (section V.C) indicate that this initially compressive strain pulse arises almost entirely from the rapid heating of the bottom electrode and propagates initially towards the surface [26,27]. The strain pulse reflected from the free surface at time $\tau=27$ ps, returned through the SL, accompanied by lattice expansion, reached the SL/SRO electrode interface at time $2\tau$, and then propagated into the substrate. At times greater than $2\tau \approx 54$ ps, the acoustic distortion in the SL is comparatively small because little of the acoustic energy is reflected at the SL/SRO interface. The optically induced lattice expansion arising from depolarization screening observed after time $2\tau$ is 0.06%. The time dependence of $\Delta Q_z$ is accurately predicted by an acoustic simulation including two sources of excitation: (i) the propagation of an acoustic pulse from SRO electrode and (ii) photoinduced expansion due to optical absorption within the SL. The stress arising from the depolarization field screening in the PTO-STO layer is assumed in the simulation to occur immediately after the optical absorption. The depolarization-field-screening effect persists for the several-nanosecond lifetime of the excited carriers, leading to the expansion at the longest times in Fig. 1(e) [34].

The structural distortion is also apparent in the intensity distribution in a region of reciprocal space including the SL 002 Bragg reflection and nearby film-thickness intensity oscillations, as shown in Fig. 2(a) for an optical fluence of 1.67 mJ cm$^{-2}$, which accurately matches the simulation, as in Fig. 2(b). Diffraction simulations considering only optical absorption in the SL are significantly different from the observations in Fig. 2(b), notably missing the reflection and sign reversal at time $\tau$, as discussed in section V.F.

### III. ULTRAFAST POLARIZATION ROTATION

The time dependence of the intensity as a function of $Q_x$ is shown in Fig. 3(a). The maxima of the intensity of the domain diffuse scattering occurred at constant $Q_x$, indicating that the domain periodicity is not changed by photoexcitation. The normalized domain diffuse scattering intensity varies significantly following optical excitation, as shown in Fig. 3(b). The normalized domain diffuse scattering intensity was lower than the initial intensity during the propagation of the compressive acoustic pulse, reaching a minimum of 0.75 at $t=18$ ps. After the reflection of the acoustic pulse from the surface, which changes the sign of the strain pulse, the normalized domain diffuse scattering intensity again decreased to 0.9 at $t=45$ ps. Finally, after the propagation of the reflected acoustic pulse into the substrate, the normalized domain diffuse scattering intensity was approximately constant at 1.05. The domain diffuse scattering intensity intuitively depends on the difference between the polarizations of the up and down domains in the striped pattern, and can be expected to increase as the average polarization difference increases [35]. The detailed simulations below show that this intuitive picture is correct, but that the specific scaling of the intensity is complicated because the strain (and hence polarization) is non-uniformly distributed during the propagation of the acoustic pulse.

The independent experimental measurements of domain diffuse scattering intensity and the structural distortion (i.e. Figs. 2(a) and 3(b)) allow the effects of strain and optical absorption on the polarization to be determined precisely. Changes in the unit cell-scale ferroelectric polarization can be included in x-ray scattering simulations by varying the Ti displacement along the out-of-plane direction within the PTO and STO unit cells. When the strain and polarization are aligned, the LGD theory predicts that $P_z$ scales with the tetragonality of the unit cell $\varepsilon_T=c/a-(c/a)_{\text{para}}$ such that $P_z \propto \varepsilon_T^{1/2}$, where $c$ and $a$ are the out-of-plane and in-plane lattice parameters, respectively [36], and $(c/a)_{para}$ is the tetragonality of SLs in the competing paraelectric phase [10]. The details of the domain diffuse scattering simulation are provided in Section V.D.

The $\varepsilon_T^{1/2}$ model cannot describe $P_z$ in ferroelectrics exhibiting a complex polarization distribution, as in the SLs, because the effect of polarization rotation is ignored. The polarization of rhombohedral ferroelectrics, for example, varies strongly as a function of distortion along the pseudocubic axes [37]. Following this expectation, an x-ray scattering simulation based on an $\varepsilon_T^{1/2}$ relationship does not accurately reproduce the experimental intensity variation, as shown in Fig. 3(b). The agreement with the $\varepsilon_T^{1/2}$ model is particularly poor during the expansion resulting from the reflection of the acoustic pulse at the surface and after the reflected acoustic pulse passes into the substrate, e.g. at time 40 ps in Fig. 3(b).

The rapid increase in $P_z$ via the polarization rotation mechanism involves a stronger dependence than expected from the uniform polarization model and can be simulated using $P_z \propto \varepsilon_T^n$, where $n$ is an exponent that we expect to be larger than 1/2. An x-ray scattering simulation including polarization rotation (section V.D) was developed using atomic positions computed by changing the fractional atomic displacement within each unit cell to produce the change in $P_z$ required by the polarization rotation model. The value of $n$ was varied to match the simulated domain diffuse scattering to the experimental time depenedence. The measured domain diffuse scattering intensity is compared to the intensity predicted by the x-ray scattering simulation in Fig. 3(b). The strain induced by depolarization field screening is treated in the simulation as an acoustic pulse traveling through the SL, driven by the stress that results from optical absorption [4]. The depolarization-field-screening-driven strain is completely developed at time $\tau$, approximately 27 ps, when this pulse reaches the bottom electrode. The best agreement between the experiment and the empirical model was obtained with larger values of the exponent $n = 1.1$ for the depolarization field screening effect and $n=1/2$ for the strain pulse arising from SRO layer, respectively. The exponent $n=1/2$ is also applied to determine the polarization change due to the lattice contraction due to the heat conduction from SRO to the SL. The only adjustable parameter in the analysis of the domain diffuse scattering is the exponent $n$ relating the polarization to the tetragonal distortion. The time-dependence of the strain profile was separately obtained by interpreting the superlattice diffraction profile.

We cannot yet attach physical significance to the specific values of $n$ for the depolarization field screening effect, except to emphasize that it is significantly larger than 1/2. In addition, the dependence of $n$ on the source of excitation is consistent with direct coupling between the depolarization-screening effect and the polarization mediated by an electronic effect and later resulting in lattice expansion.

Further insight into the contributions of the depolarization field screening-induced strain and the acoustic pulse propagating from the SRO electrode to the change in polarization can be obtained by considering the predictions of the strain propagation model in more detail. The time dependence of the average strain and the corresponding change in polarization $\Delta P_z$ in the simulation are shown in Figs. 4(a) and 4(b), respectively. The volume-average strain, as in Fig. 4(a), has a different detailed time dependence than the fitted peak wavevector derived from the same simulation, shown in Fig. 1(e). The comparison in Fig. 4 neglects the relatively small contribution at the largest values of elapsed time $t$ from the heating of the SL by thermal conduction from the SRO. A direct comparison between the polarization changes in Fig. 4(b) and the domain diffuse scattering intensity in Fig. 3(b) is not possible because the intensity during the period of the acoustic pulse is also affected by the inhomogeneity of the strain distribution. The overall trends, however, are clear: the polarization initially decreases during the compressive strain wave and then increases during the reflected wave and in the period following the acoustic pulse in which only the depolarization-field-screening effect is observed.

Further inspection of Fig. 4(b) shows that the minima and maxima of the simulated $\Delta P_z$ occur as the compressive and expansive strain waves propagate, respectively. Crucially, the predicted total $\Delta P_z$ is larger for the later times at which the strain contribution from the SRO-driven acoustic pulse is expansive because (i) at later times the depolarization-field-screening induced strain has increased and leads to a larger $\Delta P_z$ than an SRO-driven acoustic pulse with the same magnitude and (ii) the two contributions have the same sign in this time range. The contribution of the depolarization-field-screening-driven lattice expansion causes the difference in the magnitudes of the observed changes in the domain diffuse scattering intensity at $t<\tau$ and the increase at $t>\tau$.

The predicted increase in $P_z$ is consistent with the change that can be geometrically allowed due to polarization rotation. The initial and final states are shown schematically in Fig. 1(b). The rotation begins from an equilibrium polarization distribution that results from a combination of in-plane polarization at the PTO/STO interfaces and domain boundaries [11,13,38,39]. The average $P_z$ of all unit cells in the equilibrium PTO layer is 30% lower than the approximately unperturbed $P_z$ at the center of the domain [13]. The maximum change in the magnitude of $P_z$ that can arise due a reduction in the initial in-plane component of the polarization is thus on the order of 30%, which is larger than the increase by 22% in the simulation results matching the experiment. This reduction of the in-plane component of the polarization corresponds to a change in the displacement of the ions within the crystallographic unit cell and can be visualized as a rotation of the polarization in those regions towards the surface normal. In the configuration reached after optical excitation, the PTO layers lack the in-plane component arising from the equilibrium distribution and have a magnitude of $P_z$ equaling the expected PTO polarization. The period of the domain pattern is not affected by the propagation of the strain pulse.

The observed distortion of the SL can be compared with other optically driven excitation mechanisms. The displacive excitation of coherent phonons (DECP) couples optical effects to optical modes of crystals or superlattices, without a net lattice expansion or contraction [40,41]. The lowest-frequency optical modes are folded phonon modes of the SL, with oscillation frequencies on the order of 1 THz, far higher than the frequency range spanned by the intensity oscillations in Fig. 2. We can thus conclude that for the range spanned here, the DECP mechanism has a minimal role in excitation.

It is also useful to compare the observed results with the expectations based on the deformation potential. The steady-state expansion due to the deformation potential in $SrTiO_3$ with electron concentration $c_n$ is $\beta_e\, c_n$, with $\beta_e = 1.1 \times 10^{-24}$ cm$^3$ [42]. The optical excitation described above produces a carrier concentration of on the order of $1 \times 10^{19}$, consisting of both holes and electrons. The expected expansion is thus on the order of $10^{-5}$. The value of $\beta_e$ could in principle be different than reported by Janotti *et al.* [42] in the case of the excitation of equal numbers of electrons and holes rather than the impurity doping considered computationally. There is also some variation in the value of the coefficients describing the deformation potential [43]. It is clear, however, that the magnitude of the strain induced by the deformation potential effect is far smaller than the expansion of $3 \times 10^{-4}$ apparent in Figs. 1(e) and 1(f).

Other sources for the change in the domain diffuse scattering intensity can also be considered. It is possible, in principle, that the volume fractions of domains with polarization parallel to and anti-parallel to the surface normal could change as a result of the optical excitation. Domain-wall motion would result in an increase in the fraction of one polarization from the initial configuration of equal up and down polarizations. An increase in the volume fraction of either polarization direction yields a decrease in the domain diffuse scattering intensity [35]. The experimental results in Fig. 3(b), however, exhibit an increase in the diffuse scattering intensity. In addition, typical changes in the domain diffuse scattering intensity in applied electric fields occur over periods of nanoseconds, even in electric fields with very large magnitudes of on the order of hundreds of kV/cm [44]. The changes in domain diffuse scattering intensity in Fig. 3(b) occur, instead, over a period of tens of picoseconds. Beyond the polarization configuration issue, it is important to note that a separate optically induced heating of the SL would yield a compressive strain along the surface normal, opposite to the observed expansion following the propagation of the acoustic pulse.

## IV. CONCLUSION

Ultrafast optical excitation induces a dramatic change in the polarization within the nanodomain polarization of ferroelectric/dielectric superlattices. The changes occur over the picosecond timescale following ultrafast optical excitation and can be separately linked to both direct femtosecond-optical-pump induced depolarization-field-screening and the large amplitude of the strain pulse from the underlying electrode layer. The two sources of strain arising from optical excitation here have significantly different microscopic mechanisms and thus can have very different relationships between the lattice parameter and the polarization. The effect of elastic stress along the polarization direction in tetragonal ferroelectrics is readily described via the LGD theory and has exponent $n=1/2$. The depolarization-field screening effect is a far more complex phenomenon than the externally imposed elastic strain and can be thought of as a mechanical effect induced by the electronic screening.

The polarization change is physically enabled by a nanoscale rotation of the polarization, a phenomenon that is uniquely possible in ferroelectrics with a complex polarization configuration. The comparatively large response to depolarization screening is possible because the initial polarization configuration includes a significant fraction of the polarization that is not along the surface-normal direction. Depolarization-field screening removes the driving force for in-plane polarization allowing the polarization to rotate towards the surface-normal direction, resulting in the large increase observed here. Future possible applications of this concept can potentially involve the perturbation of other nanoscale ordering phenomena apparent in ferroelectric/dielectric superlattices, including the development of nanoscale octahedral rotation patterns [45]. The results expand the concept of polarization rotation to nanoscale configurations and point the way towards new control over nanoscale polarization dynamics in complex polar systems.

## V. METHODS

### A. Time resolved x-ray diffraction

X-ray diffraction experiments were performed at the x-ray scattering and spectroscopy (XSS) beamline of the PAL-XFEL, using a photon energy of 9.7 keV with a focal spot with a 10 μm full width at half maximum (FWHM) [30]. The average number of x-ray photons per pulse was $1.6 \times 10^8$. The incident x-ray fluence in each pulse was measured using a normalization photodiode detecting x-rays scattered from an Al attenuator. We avoided irreversible structural damage by using lower x-ray fluence than the threshold [46]. Diffracted x rays were detected using a multi-port charge coupled detector (MPCCD) [47]. The optical pump was focused to a spot with a 200 μm FWHM. The incident angle of the x rays was in the range of 17.5 to 18.5°, near the Bragg condition for the 002 reflection of SL. The optical beam was separated from the incident x-ray beam by an angle of 10° out of the scattering plane defined by the incident and diffracted x-ray beams. The footprints of the x-ray and optical beams on the sample surface were larger than the focal spot sizes by approximately a factor of 3 along the incident beam direction.

The femtosecond optical pump beam had a central optical wavelength 400 nm, corresponding to photon energy 3.1 eV. The optical pump photon energy is close to, but slightly below, the nominal bandgaps of both PTO (3.4 eV) and STO (3.2 eV) [48,49]. The optical absorption length $\zeta_{SL}$ of the SL was $\zeta_{SL}=1$ μm, estimated using the effective medium approximation [50]. The optical pump was approximately π polarized with respect to the sample surface. The SL surface had a measured optical reflectivity $R_{SL}=1.5\%$ at the wavelength, incident angle, and polarization of the optical pump beam. The optical absorption length in the SRO bottom electrode was $\zeta_{SRO} = 30$ nm [51]. The optical calculations neglected reflection at the SL/SRO interface.

The absorbed energy per unit volume in the SL and SRO layers is $W_{abs}(z_{rel})=(F_0/\zeta)\exp(-z_{rel}/\zeta)$, where $z_{rel}$ is the depth relative to the SL surface or SL/SRO interface, $F_0$ is the incident laser fluence transmitted through the air/SL or SL/SRO interface, and $\zeta$ is the optical absorption length for the relevant layer. The values of $F_0$ are $F_{0,SL}=(1-R_{SL}) F_{in}$ for the air SL interface and $F_{0,SRO}=(1-R_{SL}) F_{in} \exp(-d_{SL}/\zeta_{SL})$ for the SL/SRO interface, where $F_{in}$ is the incident optical fluence.

The computed depth dependence of the absorbed energy per unit volume from the optical pulse is plotted in Fig. 5. The depth-integrated absorbed optical energy is 8.7% and 48% of $F_{in}$ for the SL and SRO layers, respectively. A large fraction of the optical energy is thus absorbed in the SRO electrode.

### B. Experimental determination of longitudinal acoustic sound velocity

The intensity of the Laue fringes in Fig. 2(a) oscillates with a temporal period determined by the longitudinal acoustic phonon dispersion and a phase established by the effectively instantaneous optical excitation at $t$=0 [33]. The Fourier transform of the diffracted intensity near the PTO/STO SL 002 Bragg reflection (Fig. 6) exhibits a linear dispersion of the frequency as a function of the difference between the wavevector $Q_z$ and the wavevector of the SL 002 Bragg reflection, $Q_z$(002). The slope of the dispersion in Fig. 6 is the longitudinal acoustic phonon velocity $v$=3700 m s$^{-1}$.

### C. Simulation of acoustic strain

The elastic wave propagation of the acoustic pulse arising from optical absorption in the SRO electrode along the thickness ($z$) direction of the SL/SRO/STO heterostructure can be described by [27]:

$$\rho \frac{\partial^2 u_z(z,t)}{\partial t^2} = c_{11} \frac{\partial^2 u_z}{\partial z^2} - (c_{11} + 2c_{12})\beta \frac{\partial \Delta T(z,t)}{\partial z} \qquad (1)$$

Here $u_z$ is the mechanical displacement along $z$. The mass density $\rho$, elastic stiffness coefficients $c_{11}$ and $c_{12}$, and the linear thermal expansion coefficient $\beta$ all have material-dependent values that differ for the SL, SRO, and STO substrate, as summarized in Table 1 [52-60]. For simplicity, the STO/PTO SL is treated as a homogeneous continuum within which each parameter ($\rho$, $c_{ij}$, $\beta$, etc.) has a single effective value.

The temperature increase at depth $z$ at time $t$ due to the optical pulse is $\Delta T(z,t)$. The evolution of temperature profile $T(z,t)$ is governed by the heat conduction equation,

$$\rho c_p \frac{\partial T(z,t)}{\partial t} = \kappa \frac{\partial^2 T(z,t)}{\partial z^2}. \tag{2}$$

Here $c_p$ is the specific heat capacity and $\kappa$ is the thermal conductivity. Again, both parameters are material-dependent and have different values for the SL, SRO, and STO substrate, as in Table 1. We calculate the initial temperature distribution in the laser-heated SRO layer using $T(z, t=0) = T_0 + F_{0,SRO} \frac{1}{\zeta_{SL} c_p} e^{-z/\zeta_{SL}}$, where $T_0$ is the initial temperature. Heat transport from the SRO electrode to the SL results in an additional contribution to the strain near the SL/SRO interface that arises over a period of tens of ps. Heating results in a decrease of the SL lattice parameter because the coefficient of thermal expansion $\beta$ for the SL is negative in this temperature range, as listed in Table 1 [24]. Thermal conduction from the SRO bottom electrode leads to a small thermal contraction that is evident at the longest times probed in this experiment. The thermal conduction from the SRO layer occurs slowly in comparison with the transmission of the acoustic pulse, as has been previously observed in a Pb(Zr,Ti)O$_3$ thin film [61].

The acoustic propagation was studied numerically using a one-dimensional grid consisting of 1840 cells with a cell size of 0.5 nm. The Runge-Kutta method was used to solve the acoustic and heat transfer equations with a time step of 1 fs. A central finite difference was used to calculate spatial derivatives.

We simulated the depolarization-strain-driven acoustic strain wave propagation using an analytical model [27]:

$$\frac{\partial u_z(z,t)}{\partial z} = \alpha [e^{-\frac{z}{\zeta_{SL}}}\left(1 - \frac{1}{2}e^{-\frac{vt}{\zeta_{SL}}}\right) - \frac{1}{2}e^{-\frac{|z-vt|}{\zeta_{SL}}}\text{sgn}(z - vt)]. \quad (3)$$

Here $\alpha$ is an arbitrary parameter that matches the amplitude of the depolarization-strain-driven strain to the measured value. The acoustic wave propagates from the surface into the SL, SRO, and substrate. The depth dependence of the simulated strain for the acoustic strain pulses produced by the two absorption mechanisms is shown in Fig. 7 at 7 ps after the optical excitation pulse.

### D. Domain diffuse scattering intensity simulation

The diffuse x-ray scattering pattern was simulated using the kinematic x-ray diffraction approximation. The simulation considered one in-plane repeating unit of an idealized nanodomain pattern with dimensions $N \times M$, 242 unit cells (u.c.) along the growth direction and 300 u.c. in-plane. A piece of the simulation cell spanning a single SL repeating unit and one in-plane nanodomain period is shown in Fig. 8(a). The displacements of Ti cations from non-polar positions in the unit cell were assigned to be proportional to the polarization $P_z$ in each unit cell. In-plane displacements are not included in Fig. 8(a). The fractional atomic displacements in STO layers were set to be 40% of those in PTO layers because theoretical studies of the nanodomain polarization configuration indicate that the polarization in STO layers is 30%-50% of the polarization in PTO layers [13,62]. The simulations considered the case in which the volume fraction of up and down polarizations within each nanodomain repeating unit was 50%. With the relationship between the polarization and tetragonality given by $P_z \propto \varepsilon_T^n$, the ferroelectric polarization $P_z(z,t)$ at time $t$ at a location $z$ with out-of-plane strain $\varepsilon_{33}(z,t)$ is the sum of two terms

of the form:

$$\frac{P_z(z,t)}{P_0(z)} = \frac{[\varepsilon_{T0}+\varepsilon_{33}(z,t)]^n}{\varepsilon_{T0}{}^n}. \tag{4}$$

Terms of order $\varepsilon_T \varepsilon_{33}(z,t)$ have been neglected. Here $P_0(z)$ is the initial polarization at location $z$. Two different sources of polarization were added to determine the total polarization $P_z(z,t)$: the acoustic strain with $n=1/2$, and the depolarization-field-screening-driven strain with $n=1.1$.

The x-ray intensity distribution in the domain diffuse scattering intensity was simulated by computing the square magnitude of the lattice sum. The lattice sum was computed for layer of unit cells with index $p$ such that $F_{disp,p} = \sum_{j=1}^{M} f_j \exp(-i\vec{q} \cdot \vec{r}_{j,p})$, where $M = 300$ is the number of atoms in the unit cell, $f_j$ and $\vec{r}_{j,p}$ are the atomic scattering factor and position of the $j^{th}$ atom in layer $p$, and $\vec{q}$ is the scattering wavevector. The diffracted intensity is $\left|\sum_{p=1}^{N}(F_{disp,p})\right|^2$, where $N = 242$ is the number of unit cells along the thickness direction.

The domain diffuse scattering simulation was tested by comparing the simulated intensity with the expected static scattering patterns. The simulated distribution of intensity in the $Q_x$-$Q_z$ plane is shown in Fig. 8(b) for atomic positions corresponding to a static polarization distribution matching the room temperature polarization of the PTO-STO SL. The SL 002 Bragg reflection and the domain diffuse scattering satellites appear at $Q_x = 0$ and $\pm 0.054$ Å$^{-1}$, respectively in Fig. 8(b). The high-frequency oscillations along the $Q_x$ direction in Fig. 8(b) are a result of the lateral periodicity of the simulation cell. The dependence of the intensity on the static polarization was tested by systematically varying the $\varepsilon_T$ and using atomic positions calculated using the polarization selected using $n=1/2$. The simulated domain diffuse scattering intensity under these conditions is proportional to $\varepsilon_T$, as shown in Fig. 8(c). The proportional relationship between polarization and tetragonality is consistent with previously reported experiments and calculations [12,24]. The simulated domain diffuse scattering intensity is shown as a function of $P_z$ under the assumption that $P_z$ is proportional to $\varepsilon_T^{1/2}$. The simulated intensity in Fig. 8(d) also matches an analytical description in which the domain diffuse scattering intensity is proportional to the square of $P_z$ [35].

### E. Superlattice structural characterization

The steady-state diffraction pattern of the PTO/STO SL is shown in Fig. 9, acquired at a photon energy of 11 keV at station 7-ID of the Advanced Photon Source at Argonne National Laboratory. The wavevector range of Fig. 9 spans the 002 reflections of the STO substrate and the PTO/STO SL and several reflections with periodicity corresponding to the thickness of the SL repeating unit. The SL reflections appear at $Q_z = \frac{2\pi}{d_{avg}}(m + \frac{l}{p})$, where $m$ and $l$ are integer indices, $d_{avg}$ is the average lattice of the SL, and $p$ is the number of unit cells in one repeating unit. Fig. 9 shows the reflections at $l$ =-2, -1, 0, +1 with $m$=2.

### F. Predicted strain dynamics in the absence of optical absorption in the bottom electrode

As a complement to the consideration of the experimental sample, a time-dependent diffraction simulation was used to understand the optically induced acoustic pulse propagation in the absence of optical absorption in the SRO bottom electrode. Figure 10 shows the simulated time dependence of the diffracted intensity near the SL 002 reflection for the hypothetical case in which there is optical absorption in the SL but no absorption in other layers, e.g. for an SL grown without an SRO bottom electrode. The intensity distribution in Fig. 10 corresponds to diffraction from a thin-film layer with a single strain pulse, similar to photoacoustic excitation in metallic thin films [27]. The single strain pulse arising from the optical absorption within the SL does not match the experimental results shown in Fig. 2(a). Key differences are that the single strain pulse does not produce the temporal oscillations at $t > \tau$ and there is no reversal of the sign of the strain pulse, which occurs after reflection at the surface in the experiment. Similarly, a single strain pulse resulting from absorption in the SRO bottom electrode does not reproduce the experimental results, particularly at times $t > 2\tau$ for which the absorption in the SL produces a long-duration lattice expansion.

## ACKNOWLEDGEMENTS

The authors gratefully acknowledge support from the U.S. DOE, Basic Energy Sciences, Materials Sciences and Engineering, under contract no. DE-FG02-04ER46147. H.J.L. acknowledges support by the National Research Foundation of Korea under grant 2017R1A6A3A11030959. M. D. and M. H. Y. acknowledge support from the U. S. National Science Foundation through grant no. DMR-1334867. S. D. M. acknowledges support through the University of Wisconsin Materials Research Science and Engineering Center through NSF grant no. DMR-1720415. J.H. acknowledges support from the National Science Foundation under award CBET-2006028. The free-electron-laser experiment was performed at the XSS beamline of PAL-XFEL (proposal no. 2018-1st-XSS-003) funded by the Ministry of Science and ICT of Korea. This research used resources of the Advanced Photon Source, a U.S. Department of Energy (DOE) Office of Science User Facility, operated for the DOE Office of Science by Argonne National Laboratory under Contract No. DE-AC02-06CH11357. We are grateful for the assistance of Dr. Joonkyu Park in the acquisition of the steady-state superlattice diffraction pattern.

in PbTiO$_3$/SrTiO$_3$ Superlattices from First Principles, Phys. Rev. B **85**, 184105 (2012).

[14] S. Li, Y. J. Wang, Y. L. Zhu, Y. L. Tang, Y. Liu, J. Y. Ma, M. J. Han, B. Wu, and X. L. Ma, *Evolution of Flux-Closure Domain Arrays in Oxide Multilayers with Misfit Strain*, Acta Mater. **171**, 176 (2019).

[15] C. v. Korff Schmising, M. Bargheer, M. Kiel, N. Zhavoronkov, M. Woerner, T. Elsaesser, I. Vrejoiu, D. Hesse, and M. Alexe, *Coupled Ultrafast Lattice and Polarization Dynamics in Ferroelectric Nanolayers*, Phys. Rev. Lett. **98**, 257601 (2007).

[16] K. Istomin, V. Kotaidis, A. Plech, and Q. Kong, *Dynamics of the Laser-Induced Ferroelectric Excitation in BaTiO$_3$ Studied by X-ray Diffraction*, Appl. Phys. Lett. **90**, 022905 (2007).

[17] C. Hauf, A.-A. H. Salvador, M. Holtz, M. Woerner, and T. Elsaesser, *Soft-Mode Driven Polarity Reversal in Ferroelectrics Mapped by Ultrafast X-ray Diffraction*, Structural Dyn. **5**, 024501 (2018).

[18] R. Mankowsky, A. von Hoegen, M. Först, and A. Cavalleri, *Ultrafast Reversal of the Ferroelectric Polarization*, Phys. Rev. Lett. **118**, 197601 (2017).

[19] Y. Ahn, A. Pateras, S. D. Marks, H. Xu, T. Zhou, Z. L. Luo, Z. H. Chen, L. Chen, X. Y. Zhang, A. D. DiChiara, H. D. Wen, and P. G. Evans, *Nanosecond Optically Induced Phase Transformation in Compressively Strained BiFeO$_3$ on LaAlO$_3$*, Phys. Rev. Lett. **123**, 045703 (2019).

[20] H. J. Lee, T. Shimizu, H. Funakubo, Y. Imai, O. Sakata, S. H. Hwang, T. Y. Kim,

FIG. 1. Photoinduced picosecond dynamics in nanodomains. (a) Femtosecond optical excitation of ferroelectric nanodomains. (b) Mechanisms of polarization distortion accompanying expansion following optical excitation. (c) Slice of reciprocal space near the 002 reflection at $t < 0$. The diffuse scattering intensity at $Q_x = \pm 0.054$ Å$^{-1}$ arises from the nanodomain pattern. (d) Timeline of structural effects induced by optical excitation: acoustic propagation and reflection and depolarization-field-screening-driven strain. (e) Normalized domain diffuse scattering intensity as a function of $Q_z$ following optical excitation at $t=0$. The intensity was normalized with respect to the intensity at $t < 0$. (f) Measured (points) and simulated (line) shift $\Delta Q_z$ of the peak out-of-plane wave vector of the maximum domain diffuse scattering.

FIG. 2. Photoinduced lattice distortion of the PTO/STO SL. (a) Measured and (b) simulated time dependence of the distribution of intensity as a function of $Q_z$ near the SL 002 Bragg reflection. Intensity is normalized to the peak intensity of the 002 Bragg reflection at $t < 0$.

FIG. 3. Picosecond dynamics of polarization distortion in nanodomains. (a) Domain diffuse scattering intensity as a function of time following optical excitation and wavevector $Q_x$. (b) Time dependence of measured (squares) and simulated (lines) normalized domain diffuse scattering intensity. The solid and dashed lines represent simulations based on the enhanced polarization response with $n=1.1$, termed the polarization rotation model, and the response expected with a uniformly polarized tetragonal ferroelectric ($n=1/2$), termed the $\varepsilon_T^{1/2}$ model.

FIG. 4. Details of the polarization rotation model. (a) Simulated average out-of-plane strain in the SL, under conditions matching the experiment shown in Fig. 1(e). (b) Simulated total change in out-of-plane polarization, $\Delta P_z$, and separate contributions due to acoustic strain and depolarization field screening-induced expansion.

FIG. 5. Absorbed optical energy per unit volume in the SL/SRO thin film heterostructure.

FIG. 6. Time-domain Fourier transform of the diffracted x-ray intensity near the 002 SL Bragg reflection. The intensity was normalized to the maximum intensity. The slope of the dispersion (dashed line) is $v$=3700 m/s.

FIG. 7. Simulated acoustic strain profiles produced by depolarization-field-screening (red) and heating of the SRO bottom electrode (blue) as function of depth at 7 ps after optical excitation.

FIG. 8. (a) Atomic arrangement within a cell with a lateral size matching a single nanodomain period of one structural repeating unit of the PTO/STO SL. The solid line indicates the location of the boundary between up and down polarization regions. (b) Simulated distribution of scattered x-ray intensity in the $Q_x$-$Q_z$ plane of reciprocal space in a region including the 002 SL Bragg reflection. Domain diffuse scattering intensity maxima appear at $Q_z$=3.13 Å$^{-1}$ and $Q_x$ = ± 0.054 Å$^{-1}$. Fringes arising from the finite lateral size of the simulation cell appear with periodicity of approximately 0.007 Å$^{-1}$ along $Q_x$. (c) Simulated normalized domain diffuse scattering intensity as a function of tetragonality $\varepsilon_T$. (d) Simulated normalized domain diffuse scattering intensity as a function of polarization under conditions in which the change in the ferroelectric polarization follows $\varepsilon_T^{1/2}$. The dashed line in (d) is calculated using the model given in [35] in which the domain diffuse scattering intensity is proportional to the square of $P_z$.

FIG. 9. Diffraction pattern of the PTO/STO SL along the line $Q_x$=$Q_y$=0.

FIG. 10. Simulated diffraction profiles corresponding to the hypothetical case of a single strain pulse produced by depolarization-field screening in the PTO/STO SL, in the absence of the strain pulse arising from the photoacoustic excitation of the SRO electrode.

Table 1. Parameters for the simulation of acoustic pulse propagation and heat transport.

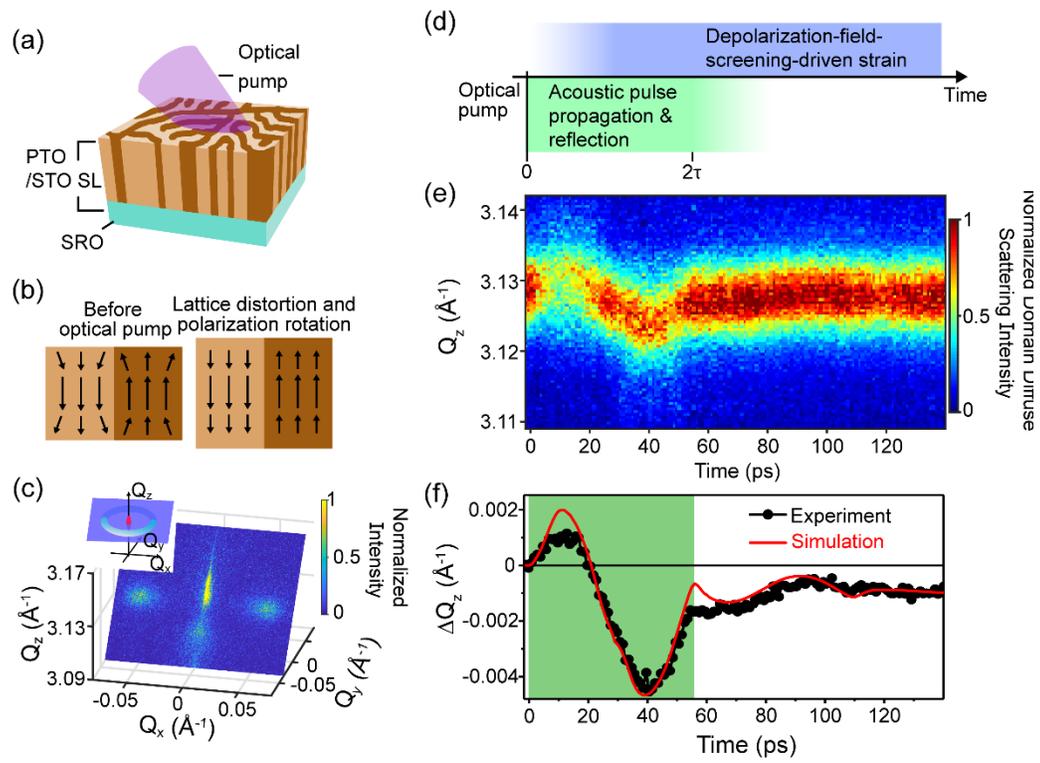

Lee *et al.,* Figure 2

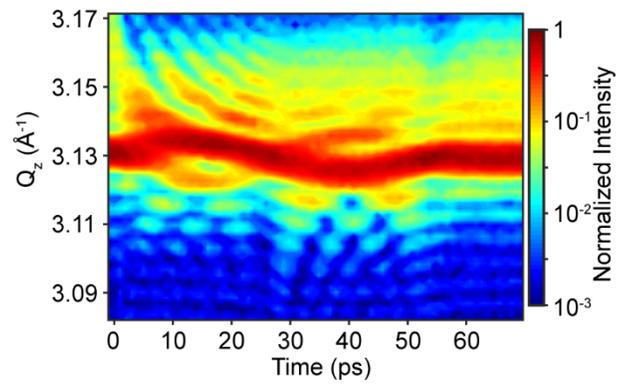

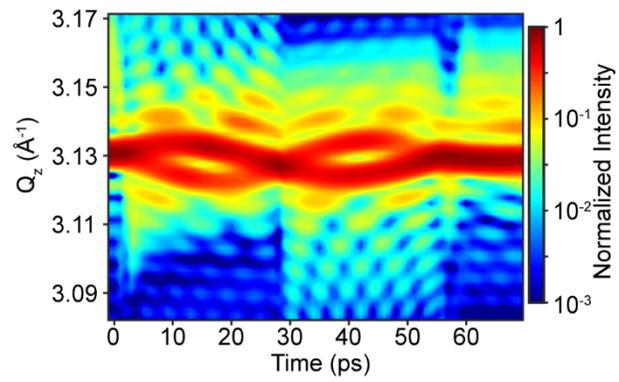

Lee *et al.,* Figure 3

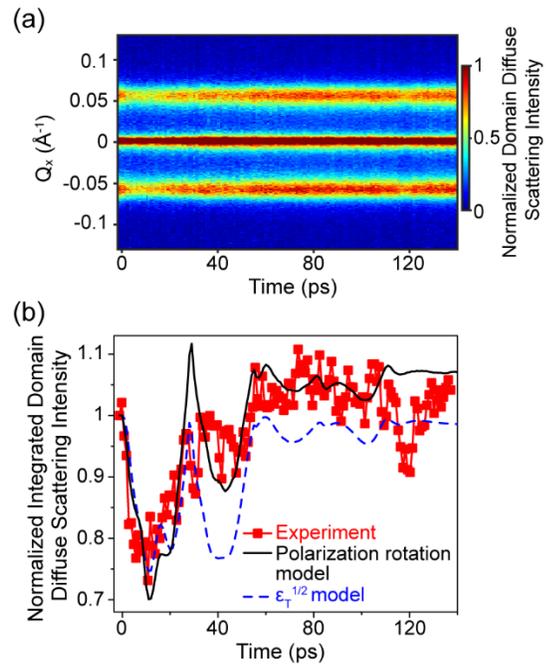

Lee *et al.,* Figure 4

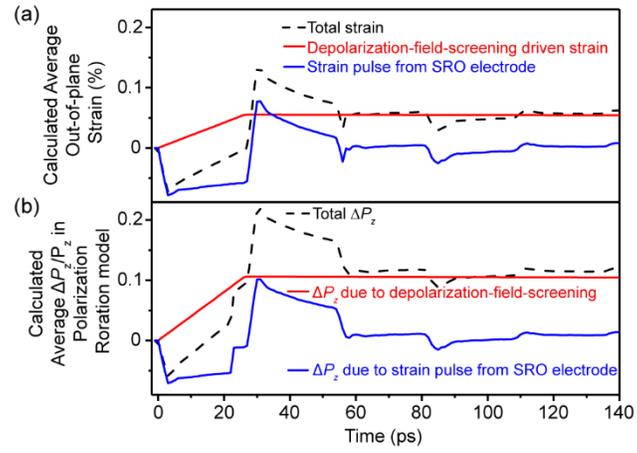

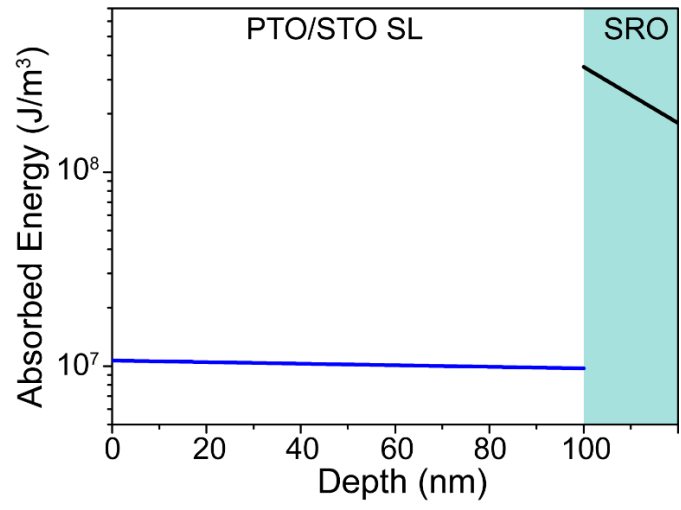

Lee et al., Figure 6

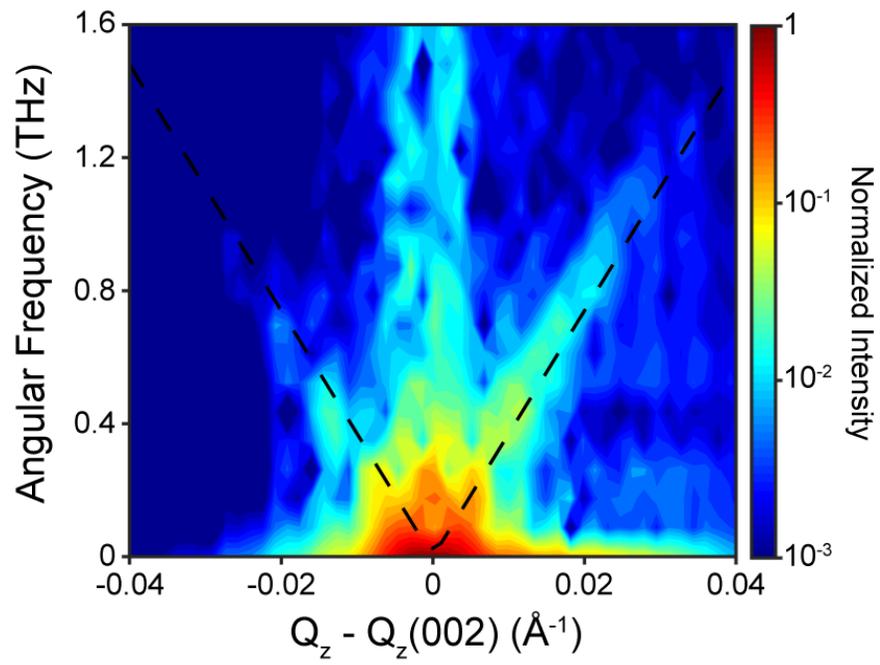

Lee *et al*., Figure 7

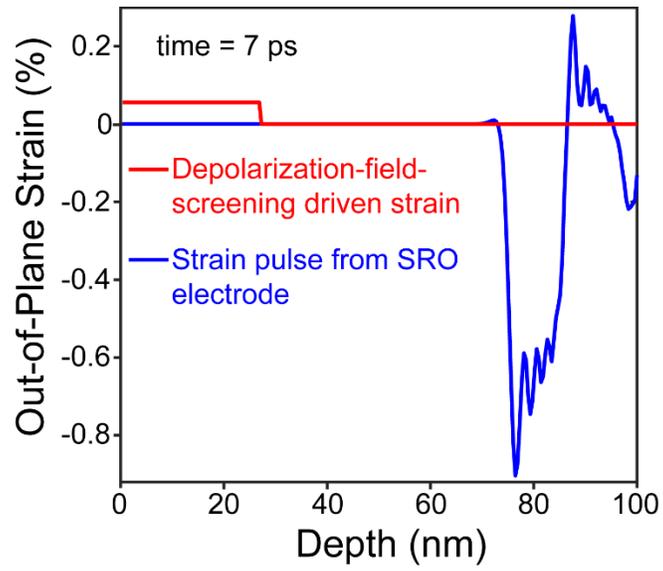

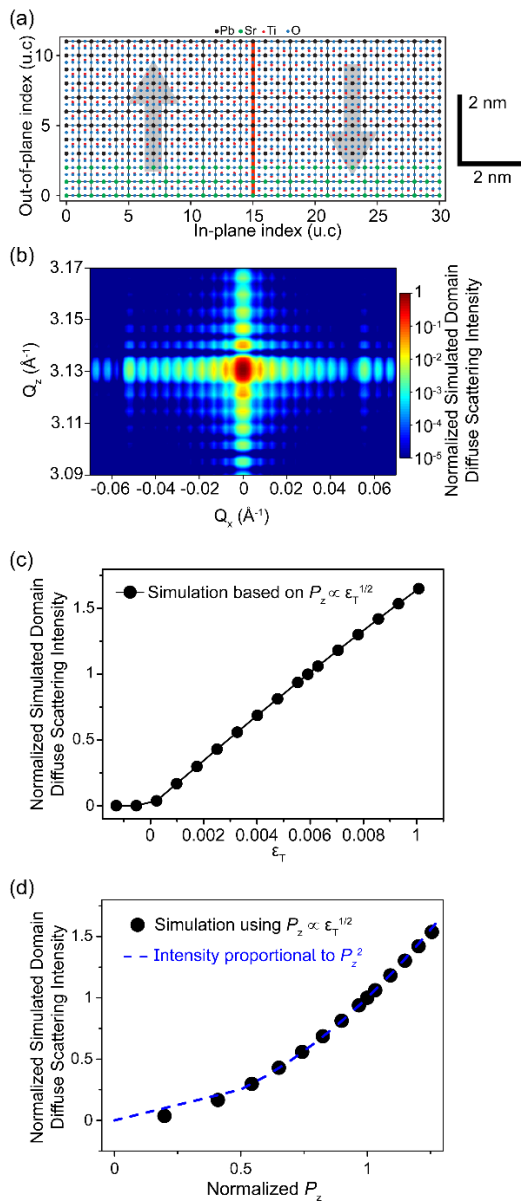

Lee *et al.*, Figure 9

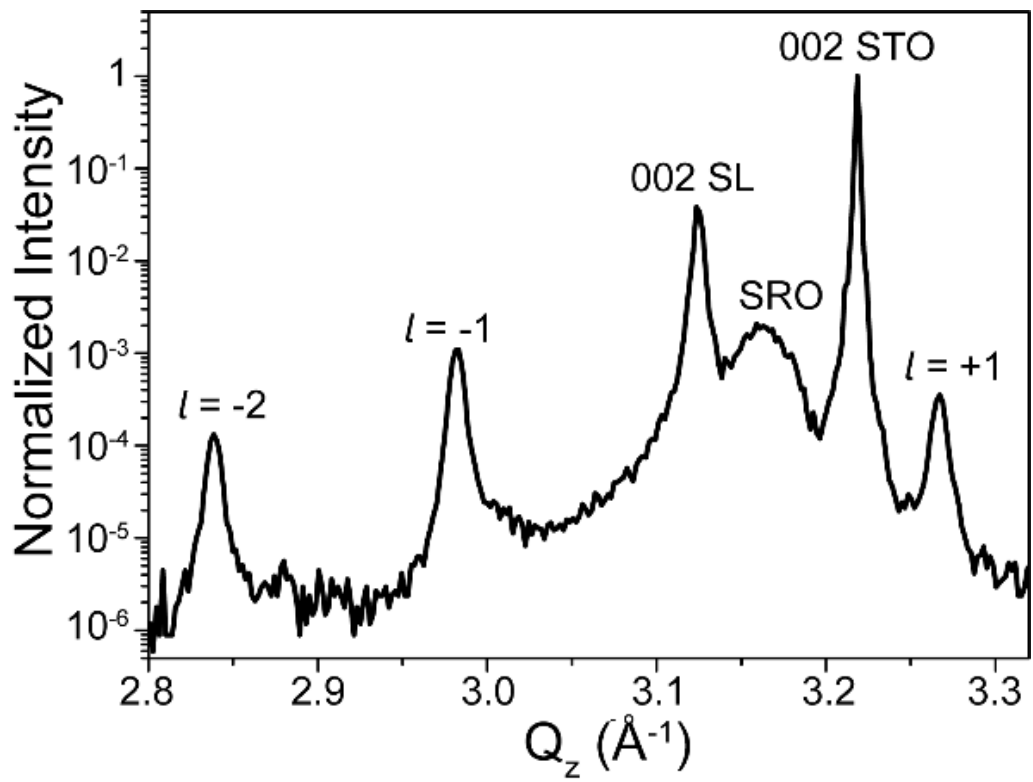

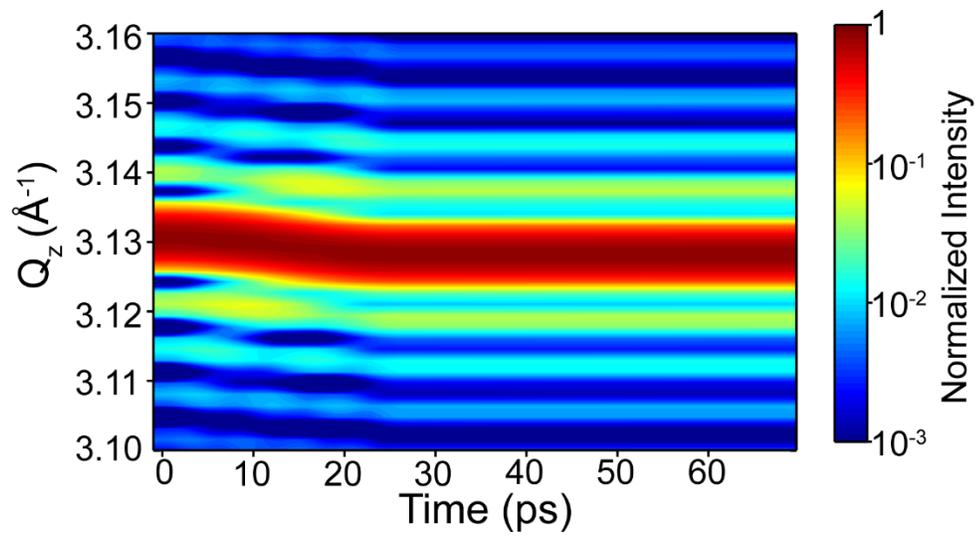

Lee *et al.*, Table 1

| Parameter | PTO | STO | SRO | SL |
|---|---|---|---|---|
| $\rho$ (kg m$^{-3}$) | 7970 | 5117 [28] | 6526 [28] | 7192 * |
| $c_{11}$ (GPa) | | 311.3 [52] | *** | 98.46 ** |
| $c_{12}$ (GPa) | | 102.4 [52] | *** | 32.82 ** |
| $\kappa$ (W m$^{-1}$ K$^{-1}$) | 5.23 [53] | 11.16 [54] | 5.97 [55] | 6.116 * |
| $c_p$ (J Kg$^{-1}$ K$^{-1}$) | 294.75 [56] | 488.568 [57] | 383.08 [55] | 347.61 * |
| $\beta$ (K$^{-1}$) | | $9 \times 10^{-6}$ [58] | $4.5 \times 10^{-5}$ [59] | $-1.44 \times 10^{-5}$ [24]**** |

* Weighted average of values for PTO and STO.
** Determined using the measured longitudinal sound velocity $v$= 3700 m/s and Poisson ratio 0.25.
*** The values of $c_{11}$ and $c_{12}$ for SRO taken to be the same as STO, following the reported similarity of values of elastic parameters for SRO and STO [60].
**** The value of $\beta$ for the optical excitation experiment was calculated by subtracting the additional lattice contraction effect arising from the thermal expansion of the substrate.